\documentclass[aps,prl,twocolumn,showpacs,floatfix,amsmath,amssymb]{revtex4}

\usepackage{amsmath,bm,amsfonts}
\usepackage{graphicx}

\begin{document}

\title{Linear scaling Krylov subspace method for large scale
      {\it ab initio} electronic structure calculations of metals}

\author{Taisuke Ozaki}
\address{
     Research Institute for Computational Sciences (RICS),
     National Institute of Advanced
     Industrial Science and
     Technology (AIST),
     1-1-1 Umezono, Tsukuba,
     Ibaraki 305-8568, Japan 
}


\date{\today}

\begin{abstract} 
 An efficient and robust linear scaling method is presented for large
 scale {\it ab initio} electronic structure calculations of a wide variety
 of materials including metals. The detailed short range and the effective
 long range contributions to the electronic structure are taken into account
 by solving an embedded cluster defined in a Krylov subspace, 
 which provides rapid convergence for not only insulators but also metals.
 As an illustration of the method, we present a large scale calculation
 based on density functional theory for a palladium cluster with
 a single iron impurity.
\end{abstract}


\maketitle

The development of linear scaling method is a promising
direction of extending applicability of {\it ab initio} electronic
structure calculations based on density functional theory
(DFT) to large scale realistic systems\cite{KS,ON1}.
In fact, over the last decade,
considerable efforts have been devoted to establish efficient
and robust linear scaling methods\cite{ON1,ON2,ON3,Ozaki1}, and 
successful applications have been reported within non-SCF
tight binding (TB) scheme\cite{Applications}.
Nevertheless, within the fully self consistent field (SCF) DFT
much improvement is still needed to reduce the error below chemical
accuracy (a few milli-Hartree/atom), referred to as milli-Hartree
accuracy hereafter, for a wide variety of materials with modest
computational cost.
In addition, the application to {\it ab initio} calculations is
practically hampered by an intractable feature that an approximate
solution of eigenstates by the linear scaling methods often induces
instabilities in the SCF calculation.
In this paper to overcome these difficulties we present an efficient
and robust linear scaling method for a wide variety of materials
including metals in which ideas behind two linear scaling
methods, divide-conquer (DC)\cite{ON2}
and recursion methods\cite{ON3,Ozaki1},
are unified in a single framework.
It is known that the DC method provides rapid
convergence for covalent systems such as biological molecules
with numerical stability during the SCF calculation\cite{ON2}.
However, the application of the DC method to metals is significantly
restricted by requirement of the large size of truncated cluster.
On the other hand, the recursion method based on Lanczos
algorithms and Green's functions is one of
suitable methods for metals\cite{ON3,Ozaki1},
although the SCF calculation with 
the recursion method becomes unstable. The main idea behind the
recursion method is to employ a Krylov subspace generated by 
the Lanczos algorithm in evaluating Green's functions\cite{ON3,Ozaki1,Lanczos},
and this is the reason why the recursion method can provide
rapid convergence for metals. 
Thus, we propose a novel method which possesses the advantages in
two methods and overcomes the
drawbacks.
Let us assume that the basis set consists of nonorthogonal
localized functions such as pseudo-atomic orbitals (PAO)\cite{Ozaki2}
and finite elements basis\cite{Tsuchida}.
Throughout this paper we use the PAO $\chi$ as basis
function to expand one-particle wave functions\cite{Ozaki2}.
The charge density $\rho^{\sigma}({\bf r})$ associated
with spin component $\sigma$ is evaluated via Green's
function $G$ by 
\begin{eqnarray}
  \rho^{(\sigma)}({\bf r})
     = \sum_{i\alpha,j\beta}
       \chi_{i\alpha}({\bf r})\chi_{j\beta}({\bf r})
       n^{(\sigma)}_{i\alpha,j\beta},
\end{eqnarray}
where $i$ is a site index, $\alpha$ an orbital index, and
$n^{\sigma}_{i\alpha,j\beta}$ density matrix given by 
\begin{eqnarray}
  n^{(\sigma)}_{i\alpha,j\beta} = -\frac{1}{\pi}
             {\rm Im}\int G^{(\sigma)}_{i\alpha,j\beta}(E+{\rm i0^+})
             f(\frac{E-\mu}{k_BT})dE
\end{eqnarray}
with the Fermi function $f(x)\equiv 1/[1+\exp(x)]$.
Since only the charge density $\rho^{\sigma}$ and the density
matrix $n^{\sigma}$ are required in the conventional DFT, 
we focus on the evaluation of Green's functions in
later discussion. The spin index $\sigma$ will hereafter
be dropped for simplicity of notation.
It is noted that the DC and recursion methods provide
different ways of evaluating Green's functions from
a local Hamiltonian $H$ and overlap $S$ matrices
constructed from the local environment for each atom\cite{ON2,ON3,Ozaki1}.
By taking into account the rapid convergence as observed
in the recursion method and considering that the Krylov
subspace for the non-orthogonal basis functions is defined by 
$\{|W_0), S^{-1}H|W_0), (S^{-1}H)^2|W_0),
..., (S^{-1}H)^N|W_0)\}$, we introduce
a Krylov subspace ${\bf U}$ for each atom given by 
\begin{eqnarray}
 {\bf U}={\bf W}{\bf X}\lambda^{-1},
\end{eqnarray}
where 
${\bf W}\equiv \{|W_0),|W_1),|W_2),...,|W_N)\}$,
$\lambda$ and ${\bf X}$ are eigenvalues and corresponding 
eigenvectors of an overlap matrix ${\bf W}^{\dag} {\hat S}{\bf W}$.
The {\bf W} is generated by the following procedure:
(i)   $|R_n) = QH|W_{n-1})$,
(ii)  $|W'_n) = |R_n) - \sum_{m=0}^{n-1}|W_m)(W_m|\hat{S}|R_m)$,
(iii) $|W_n)=$ $S$-orthonormalized block vector of $|W'_n)$.
In this procedure, $Q$ is the inverse of a local overlap
matrix $S$ constructed from the same truncated cluster
as in the construction of the local Hamiltonian $H$, where
the cluster is constructed by a logically truncation method\cite{Ozaki1}.
In case the large size of truncated cluster is required for
reduction of the computational error, $Q$ can be substituted
by an approximate inverse given by ${\bf V}{\bf s}^{-1}{\bf V}^{\dag}$ with
${\bf s}={\bf V}^{\dag}\hat{S}{\bf V}$ and 
${\bf V}\equiv\{|V_0),|V_1), |V_2),..., |V_{N'})\}$
which is generated by the following procedure:
(I)   $|Y_m) = S|V_{m-1})$,
(II)  $|V'_m) = |Y_m) - \sum_{p=0}^{m-1}|V_p)(V_p|Y_m)$
(III) $|V_m)=$ I-orthonormalized block vector of $|V'_m)$.
The initial states $|V_0)$ and $|W_0)$ consist of block
I-orthonormalized vectors and its $S$-orthonormalized vectors,
and the optimum choice of $|V_0)$ depends on the system
as discussed later on.
In the generation scheme, we impose only the $S$-orthogonality between
Krylov vectors without assuming any specific form for the representation
of the Hamiltonian matrix, while a tridiagonal form of the Hamiltonian matrix
is imposed in the Lanczos algorithm\cite{ON3,Ozaki1}.
Although the procedure (i)-(iii) gives a set of $S$-orthonormal
Krylov vectors in principle, the $S$-orthonormality is not well assured
due to round-off error in the Gram-Schmidt orthogonalization.
Therefore, the Krylov subspace ${\bf U}$ is given by
an orthogonal transformation Eq.~(3). For numerical stability, 
it is crucial to construct the Krylov subspace ${\bf U}$ at
the first SCF step, and to fix it during subsequent steps. 
If the Krylov subspace is regenerated at every SCF step, the
SCF convergence becomes significantly worse because of
fluctuation of the spanned space, which is the reason for
the instability inherent in the recursion method coupled
with the SCF calculation as discussed later on.
By considering a further spatial division of the truncated cluster
into a core and the remaining buffer regions, and taking the Krylov
subspace representation, the original generalized 
eigenvalue problem $Hc_{\mu} = \varepsilon_{\mu} Sc_{\mu}$
for the truncated cluster can be transformed to a standard eigenvalue
problem $H^{\rm K} b_{\mu} = \varepsilon_{\mu} b_{\mu}$
with
\begin{eqnarray}
  \nonumber
  H^{\rm K}
  & = & {\bf U}^{\dag}H{\bf U} \\
   \nonumber
  & = & u_{\rm c}^{\dag}H_{\rm c} u_{\rm c} 
      + u_{\rm c}^{\dag}H_{\rm cb}u_{\rm b}
      + u_{\rm b}^{\dag}H_{\rm cb}^{\dag}u_{\rm c}
      + u_{\rm b}^{\dag}H_{\rm b}u_{\rm b} \\
  & = &
      H_{\rm s}^{\rm K} + H_{\rm l}^{\rm K},
\end{eqnarray}
where $H_{\rm c}$, $H_{\rm b}$, and $H_{\rm cb}$ are Hamiltonian
matrices, represented by the original basis functions $\chi$,
for the core and buffer regions, and between
the core and buffer regions, respectively.
Considering that the Krylov subspace ${\bf U}$ is decomposed to
contributions of the core and buffer regions: 
${\bf U}^{\dag}=(u_{\rm c}^{\dag},u_{\rm b}^{\dag})$,
it is straightforward to see that $H^{\rm K}$ is composed
by a short range 
$H_{\rm s}^{\rm K}\equiv u_{\rm c}^{\dag}H_{\rm c} u_{\rm c}$
and the other long range contributions $H_{\rm l}^{\rm K}$. 
\begin{figure}[t]
  \centering
  \includegraphics[width=6.5cm]{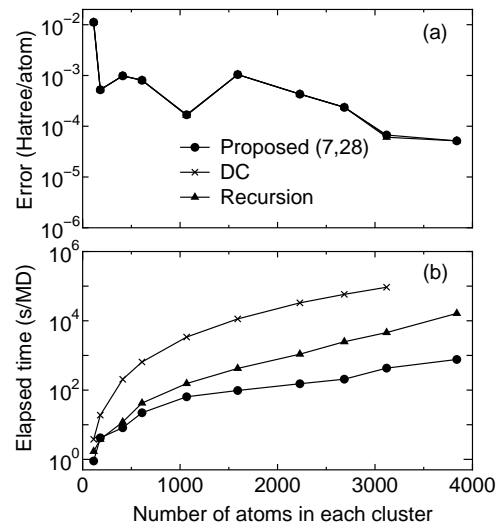}
  \caption{
     (a) absolute error, with respect to the band calculation, 
     in the total energy (Hartree/atom) for bcc lithium bulk
     as a function of atoms in each truncated cluster
     calculated by three linear scaling methods,
     (b) computational time (s) for the diagonalization part per MD step. 
     The number of atoms in the core region is 113.
     The set of numbers in the parenthesis means the percentage of the dimension
     of the subspaces ${\bf W}$ and ${\bf V}$ relative to the total
     number of basis functions in the truncated cluster, respectively.
    }
\end{figure}
Since the required buffer size to satisfy the milli-Hartree accuracy
can be large in most cases for metals, therefore, once the long
range contributions $H_{\rm l}^{\rm K}$ is calculated at the
first SCF step, the matrix is fixed during subsequent steps,
while it it possible to update $H_{\rm s}^{\rm K}$ after
achieving the self consistency. Then, the standard eigenvalue problem
is diagonalized with a updated $H_{\rm s}^{\rm K}$ and
the fixed $H_{\rm l}^{\rm K}$ during subsequent steps,
which means that the detailed short range contribution
to the electronic structure can be taken into account
with an effective correction by the long range
contribution $H_{\rm l}^{\rm K}$. Thus, the evaluation of Green's functions
is mapped to a cluster problem analogous to the
DC method\cite{ON2} but with the effective smaller Hamiltonian $H^{\rm K}$.
The core region in this study is defined by a cluster within
atoms having non zero overlap $\chi_{i\alpha}\chi_{j\beta}$ 
to the central atom $i$. In this case the required components
in the eigenvectors for the evaluation of the charge density
Eq.~(1) is easily evaluated by a back transform
$c_{\mu}=u_{\rm c}b_{\mu}$. It is noted that the evaluation
of Green's function and the density matrix Eq.~(2)
is trivial in the same way as the DC method\cite{ON2}
since we have the eigenvalues $\varepsilon_{\mu}$ and its
corresponding eigenvectors $c_{\mu}$.

\begin{figure}[t]
  \centering
  \includegraphics[width=6.5cm]{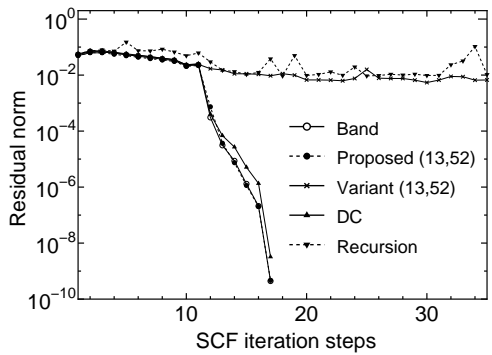}
  \caption{
    The residual norm of charge density as a function of SCF steps
    calculated by the band method, the proposed, its variant with
    the regeneration of the Krylov subspace, DC, and recursion methods
    for fcc aluminum bulk.
    In the proposed method, the core and buffer regions contain
    55 and 326 atoms, respectively.
    For the set of numbers in the parenthesis, see the caption of Fig.~1.
   }
\end{figure}

In Fig.~1 the absolute error in the total energy and the computational
time calculated by three linear scaling methods, the proposed, 
DC, and recursion methods, are shown as a function of number
of atoms in each truncated cluster for bcc lithium bulk.
All the calculations in this study were performed by a DFT code,
OpenMX\cite{Ozaki2,OpenMX}, with a generalized gradient
approximation (GGA)\cite{GGA} to the exchange-correlation potential.
It is found that three methods are equivalent in terms of 
the accuracy.
However, we see that the computational time of the proposed
method is remarkably reduced compared to those of the
DC and recursion methods.
In the proposed method the dimension of the Krylov
subspace ${\bf W}$ and that of the subspace ${\bf V}$ for the
approximate inverse of the overlap matrix
are 7~\% and 28~\% of the total number of basis functions
in the truncated cluster, respectively. In spite of the
considerable reduction of the spanned space, the method gives
the same result as that of the DC method, which clearly shows
rapid convergence of the proposed method based on the
Krylov subspace. The difference between the proposed and
recursion methods in the computational time is attributed
to the regeneration of the Krylov subspace and the evaluation
of Eq.~(2) in the recursion method.

To compare the numerical stability, the SCF convergence is
shown in Fig.~2 for the conventional band and four linear
scaling methods for fcc aluminum. 
The residual norm of charge density by the band, proposed, 
and DC methods quickly decreases, while the convergent result
is hardly obtained in the proposed method with the regeneration
of the Krylov subspace and the recursion method.
The comparison between the proposed method and its variant with
the regeneration of the Krylov subspace suggests a reason why
the recursion method tends to suffer from the numerical instability.
The regeneration of the Krylov subspace makes
the spanned subspace fluctuate, which means that an eigenvalue
problem defined by a different subspace is solved at every SCF
step. This fluctuation of the spanned space causes the difficulty
in obtaining the SCF convergence for the recursion method.
On the other hand, in the proposed method without the
regeneration of the Krylov subspace the difficulty is avoided
since the spanned space is fixed during the SCF calculation.

\begin{figure}[t]
  \centering
  \includegraphics[width=6.5cm]{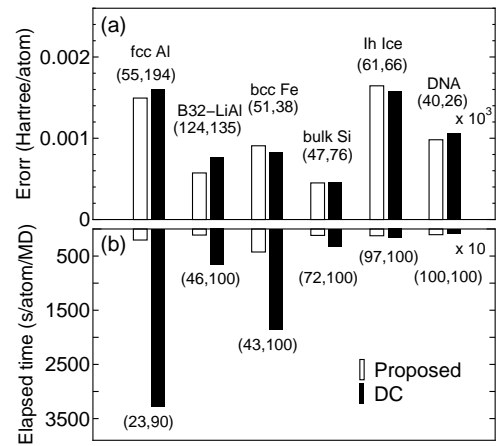}
  \caption{
     (a) absolute error, with respect to the band calculations, 
     in the total energy (Hartree/atom) calculated
     by the proposed and DC methods for metals and finite
     gap systems, (b) computational time (s/atom/MD).
     For a substantial comparison, the calculations were
     performed using a single Xeon processor.
     The set of numbers in the parenthesis of (a) means the average
     number of atoms in the core and buffer regions. For that in (b),
     see the caption of Fig.~1.
     }
\end{figure}

In Fig.~3 the absolute error in the total energy calculated by the
proposed and DC methods are shown for a wide variety of materials.
Several trends in the convergence properties can be found
in this comparison. 
It is obvious that the large truncated cluster is required to
satisfy the milli-Hartree accuracy for simple metals such as
aluminum and lithium (see also Fig.~1). However, a relatively
smaller dimension of the Krylov subspace is enough for the
convergence. For the B32-LiAl alloy and the transition metal Fe,
the relative dimension of the Krylov subspace required for
the convergence increases compared to that for the simple metals. 
For systems with a finite gap, the total energy converges to
the milli-Hartree accuracy even in a small truncated
cluster especially for DNA with a periodic double helix structure (650 atoms/unit)
consisting of cytosines and guanines, while the dimension of the Krylov
subspace for the convergence is comparable to that of the original space
defined by the truncated cluster.
Therefore, in comparison with the DC method, the proposed
method is more efficient especially for metallic systems,
and that the efficiency becomes comparable as the covalency
and ionicity in the electronic structure increase. 
The crossing point between the proposed and conventional methods
in the computational time is estimated to be about 800 atoms for
silicon bulk on the serial computation,
while it varies depending on the system and calculation conditions.
It is also interesting that the convergence rate with respect
to the Krylov subspace dimension depends on the choice of $|V_0)$. 
For metals and insulators, we find that an optimum choice 
for $|V_0)$ is a set of the basis functions in the central atom
and the central atom plus the neighboring atoms, respectively,
which may be related to different convergence properties of
constituents such as itinerant, $\sigma$, and $\pi$ electrons
in the electronic structures\cite{Ozaki1}.

\begin{figure}[t]
  \centering
  \includegraphics[width=6.5cm]{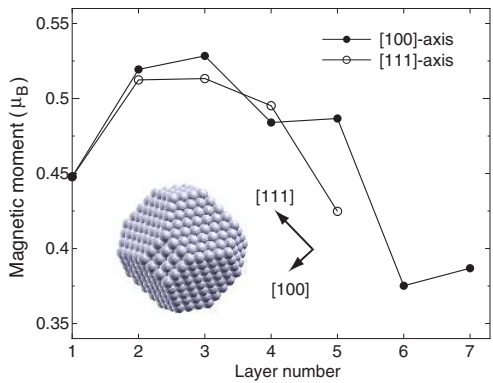}
  \caption{
      Magnetic moment of Pd atoms in a cluster, Fe$_{1}$Pd$_{806}$,
      with a central iron impurity as a function of the layer number
      along [100]- and [111]-axes. The total magnetic moment of the
      cluster is 344 ($\mu_{\rm B}$). The calculation was performed 
      using six nodes (16 CPUs/node) of a SR11K machine.
     }
\end{figure}

As an illustration of the proposed method, we present a large scale
calculation for a truncated octahedral palladium cluster with
a single iron impurity, Fe$_{1}$Pd$_{806}$\cite{appl}.
In the linear scaling calculation the truncated cluster is constructed
by atoms within a sphere with a radius of 9.2~\AA, and for example
in fcc Pd the core and buffer regions contain 55 and 170 atoms
under this condition, respectively.
The dimension of the Krylov subspace ${\bf W}$ is 600,
and the value corresponds to about 30~\% of the total number of
basis functions in the truncated cluster for fcc Pd. The exact
inverse of the local overlap matrix is used for $Q$. 
The calculation condition gives 0.00217 and 0.00012 (Hartree/atom)
as the absolute error in the total energy for fcc Pd and bcc Fe,
respectively, compared to the conventional band calculations. 
In Fig.~4 the magnetic moment of Pd atoms in the cluster Fe$_{1}$Pd$_{806}$
is shown as a function of the layer number along [100]- and [111]-axes,
where in each layer the value of the nearest atom to the axes is considered. 
It is found that a peak appears around the second or third layer 
along both the axes, and that the surface magnetic moment is smaller
than that of the inner Pd atoms. The magnetic moment of the central
iron atom is 3.68 ($\mu_{\rm B}$) which is well compared to other
theoretical and experimental values, 3.47\cite{Oswald} and
$\sim$ 4\cite{Ododo} ($\mu_{\rm B}$), in FePd alloy. 
The spatial distribution of magnetic moment may be attributed to
interactions between the inner and surface magnetic moments.
The details will be discussed elsewhere.

In summary, an efficient and robust linear scaling method has
been developed for a wide variety of materials including metals.
Based on the Krylov subspace an embedded cluster problem is solved 
with an effective Hamiltonian consisting of the detailed short range
and the effective long range contributions.
The method is regarded as a unified approach connecting
the DC and recursion methods, and enables us to obtain convergent
results with the milli-Hartree accuracy for a wide variety
of materials. The application to a palladium cluster
with a single iron impurity clearly shows that the method is a promising
approach for realization of linear scaling {\it ab initio} calculations
for metals. 

The author would like to thank Prof. K. Terakura for his continuous
encouragement.
The author was partly supported by CREST-JST and NAREGI
Nanoscience Project, the Ministry of Education, Science,
Sports, and Culture, Japan.


\begin{thebibliography}{99}
\bibitem{appl}
In these norm conserving pseudopotentials, 3d, 4s, and 4p
for iron, and 4d, 5s, and 5p for palladium were taken into account.
The basis functions, Fe5.0-s1p1d1 and Pd5.5-s1p1d1, were generated
by the orbital optimization method for bcc Fe and fcc Pd\cite{Ozaki2}. 
These pseudopotentials and basis functions give 0.0 (0.0), 0.0 (0.0)
as the magnetic moment $(\mu_{\rm B})$ of cuboctahedron Pd$_{13}$, 
and 3.86 (3.99), 0.32 (0.33) as those of cuboctahedron Fe$_{1}$Pd$_{12}$
with a central impurity Fe, where the values in the parentheses are
other theoretical results\cite{Reddy}.
\end{thebibliography}
\end{document}